\documentclass[prd,twocolumn,eqsecnum,floatfix,superscriptaddress,preprintnumbers,amsmath,amssymb,nofootinbib,aps]{revtex4-2}

\usepackage{graphicx}
\usepackage{dcolumn}
\usepackage{bm}
\usepackage{amsmath}
\usepackage{xcolor}
\usepackage{soul}
\usepackage{graphicx}
\usepackage{dcolumn}
\usepackage{bm}
\def\lesssim{\ \raise.3ex\hbox{$<$}\kern-0.8em\lower.7ex\hbox{$\sim$}\ }
\def\gesim{\ \raise.3ex\hbox{$>$}\kern-0.8em\lower.7ex\hbox{$\sim$}\ }

\begin{document}
\preprint{RIKEN-iTHEMS-Report-25}
\title{Schwinger-Keldysh approach to tunneling transport at a hadron-quark interface}
\author{Tingyu Zhang}
\affiliation{Department of Physics, Graduate School of Science, The University of Tokyo,
    Tokyo 113-0033, Japan}
\affiliation{Interdisciplinary Theoretical and Mathematical Sciences Center (iTHEMS), RIKEN, Wako 351-0198, Japan}

\author{Hiroyuki Tajima}
\affiliation{Department of Physics, Graduate School of Science, The University of Tokyo,
    Tokyo 113-0033, Japan}
\affiliation{RIKEN Nishina Center, Wako 351-0198, Japan}

\author{Motoi Tachibana}
\affiliation{Department of Physics, Saga University, Saga 840-8502, Japan}
\affiliation{Center for Theoretical Physics, Khazar University, 41 Mehseti Street, Baku, AZ1096, Azerbaijan}

\date{\today}
\begin{abstract}
We theoretically discuss quantum tunneling transport and frictions at a hadron-quark matter interface based on the Schwinger-Keldysh approach combined with the tunneling Hamiltonian, which has been developed in the context of condensed matter physics.
In the inner core of massive neutron stars, 
it is expected that cold quark matter appears at sufficiently high densities and hence exhibits color superconductivity, surrounded by nucleon superfluids at lower densities.
The perturbative expressions of the tunneling current and the friction at the interface are obtained in terms of the non-equilibrium Green's functions.
We demonstrate the DC Josephson current that occurs at the hadron-quark superfluid interface in the present scheme. 
Our framework can be applied to various conflagrations involving the interfaces relevant to astrophysical phenomena.
\end{abstract}

\maketitle

\section{Introduction}
A transport phenomenon at dense matter is an important issue to understand structures and dynamics of neutron stars, where it is expected that dense quark matter may appear at the inner core region surrounded by baryon matter~\cite{baym2018hadrons}. 
Moreover, the extremely large density leads to a baryon superfluidity~\cite{RevModPhys.75.607,haskell2018superfluidity} as well as a quark pairing state called color superconductivity~\cite{barrois1977superconducting,barrois1979non,frautschi1980asymptotic} (see also review~\cite{RevModPhys.80.1455}) that have been discussed in the connection with astrophysical observations such as cooling curves and pulser glitches in neutron stars~\cite{Blaschke_2000,lattimer2004physics}.

In superfluidity and superconductivity,
it is known that several non-trivial transport phenomena arise at the interface.
A representative example is the Josephson effect~\cite{josephson1962possible}, where the tunneling current occurs even without the external bias if the phase of the superfluid/superconducting order parameters are different in two bulk systems forming a junction.
Recently, the Josephson effect between $s$- and $p$-wave nucleon superfluids and its impact on the neutron star's rotation rates have been discussed~\cite{PhysRevD.111.023044}. 
Moreover, if the fermionic matter in the normal phase is connected with superfluid/superconducting system,
the Andreev reflection is known to occur~\cite{osti_4071988}, where
an incident particle in the normal-phase side enters the superfluid/superconducting side with forming a Cooper pair accompanying a reflected hole.
These phenomena have been studied extensively in condensed-matter~\cite{RevModPhys.80.1337} and cold-atomic systems~\cite{krinner2017two}.
Moreover,
in high-energy physics,
the Andreev reflection at the interface of cold quark-gluon-plasma and the color-flavor-locked (CFL) phase and that of the CFL phase and 
the two-flavor color-superconducting (2SC) phase
have been studied theoretically~\cite{PhysRevD.66.045024,sadzikowski2002andreev}.
Another fascinating aspect of the Andreev reflection is the analogy with the Hawking radiation from a black hole, which manifests the information mirror process at an ideal interface~\cite{PhysRevD.96.124011,PhysRevD.102.064028,zhang2023dominant}.

On the other hand,
there is
a crucial difference between condensed matter systems and dense matter in neutron stars, that is,
the confinement-deconfinement transition of hadrons at high densities governed by quantum chromodynamics (QCD)~\cite{fukushima2010phase}.
In this regard, there is a possibility of the formation of a hadron-quark interface~\cite{PhysRevD.64.074017,alford2016characteristics} between two different superfluid/superconducting phases consisting of different fermions such as baryon superfluidity and color superconductivity.
Moreover, the color degree of freedom leads to various pairing patterns in the color superconductivity~\cite{bailin1979superfluid,bailin1984superfluidity}.
These two different phases of baryons and quarks have been investigated independently at low and high density regimes,
but it is an interesting problem how these phases coexist and behave in a massive neutron star.
If baryon and quark superfluid systems have an interface, it is reasonable to expect the occurrence of the Josephson effect and the Andreev reflection across the interface.

Furthermore, in contrast to conventional condensed-matter systems in laboratories,
the tunneling transport may be induced by the macroscopic differential rotations in neutron stars.
In such a case, the velocity difference between two bulk systems may be regarded as a bias inducing the tunneling current as discussed in spintronics studies ~\cite{PhysRevB.105.L020302,PhysRevB.111.L060403}.
In particular,
the mutual friction induced by the differential rotation is relevant for the hydrodynamics of neutron star matter~\cite{baym1969spin}.

However, the hadron formation makes the analysis of the transport phenomena challenging.
In contrast to the normal-superconducting interface consisting of common fermions (e.g., electrons),
the tunneling and reflected particles may be not only quarks but also
hadrons such as baryons and mesons~\cite{juzaki2020andreev}.
This is because the baryon formation and the particle-anti-particle creation and annihilation can occur in relativistic quark systems and hence the application of the conventional Bogoliubov-de-Genne (BdG) approach based on the quasiparticle wave functions~\cite{zhu2016bogoliubov} is not straightforward.
In this sense, the formulation of the transport phenomena based on the quantum field theory is required.

To this end, the non-equilibrium Green's function approach developed by Schwinger and Keldysh~\cite{schwinger1961brownian,Keldysh}, and also described by the equations of motion in the Baym-Kadanoff approach~\cite{PhysRev.124.287,PhysRev.127.1391}, is promising.
This field-theoretical approach enables us to write down the tunneling current occurring at the interface involving bulk superfluid/superconducting systems in terms of the non-equilibrium Green's function with the so-called Keldysh contour.
It has been successfully applied to the superconducting systems, and the Josephson current and the Andreev reflection can be studied in this framework~\cite{zagoskin1998quantum,PhysRevB.54.7366,PhysRevB.71.024517}.
Moreover, it can also be applied to the tunneling transport in strongly-interacting Fermi gases with multi-particle tunneling events~\cite{PhysRevLett.118.105303,PhysRevA.95.013623,furutani2020strong,PhysRevA.106.033310,tajima2023nonequilibrium,PhysRevB.110.064512}.

\begin{figure}[t]
    \centering
    \includegraphics[width=\linewidth]{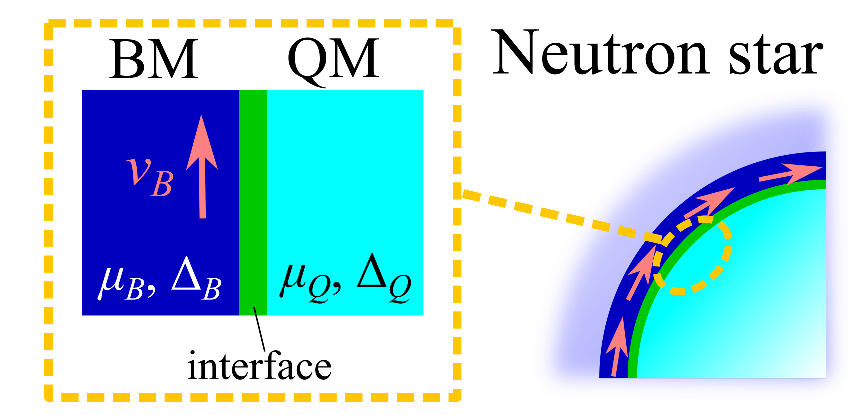}
    \caption{Schematics of the hadron-quark interface in a neutron star consisting of baryon matter (BM) and quark matter (QM). We consider the tunneling transport and the friction driven by the thermodynamic bias (e.g., difference between chemical potentials $\mu_{B/Q}$ or pairing gaps $\Delta_{B/Q}$ in BM and QM) and the BM velocity $\bm{v}_{B}$  in the rest frame of QM (i.e., the frame with vanishing QM velocity $\bm{v}_{Q}=\bm{0}$). In this sense, $\bm{v}_{B}$ can be regarded as the relative velocity between BM and QM.
    In the local area of the hadron-quark interface, the global differential rotation is considered as the linear relative motion parallel to the interface.
    }
    \label{fig:interface}
\end{figure}

In this paper,
we study the tunneling transport at the hadron-quark interface schematically depicted in Fig.~\ref{fig:interface},
within the Schwinger-Keldysh approach.
In the hybrid hadron-quark matter model with the tunneling Hamiltonian for the baryon--three-quark conversion,
we show the perturbative expression of the tunneling particle current across the interface.
Moreover, we also show that the mutual friction between hadron and quark matter can be addressed in the present approach.

This paper is organized as follows.
In Sec.~\ref{sec:model},
we present the theoretical model for the hadron-quark matter with the tunneling effect at the interfaces.
In Sec.~\ref{sec:formalism},
we show how to apply the Schwinger-Keldysh approach to the tunneling transport at the hadron-quark interface.
Moreover, we demonstrate the Josepshon current across the interace of the  baryon superfluid and the color superconductivity.
In Sec.~\ref{sec:summary},
we give a summary of this paper.

\section{Hadron-quark matter model}
\label{sec:model}
In the following, we take the unit of $\hbar=k_{\rm B}=1$.
Additionally, the bulk subsystems of baryon matter and quark matter are assumed to be in the thermodynamic limit, and their volumes are taken to be unity. 
We consider
the hybrid model consisting of relativistic baryons and quarks, where the Hamiltonian of relativistic fermions is given by~\cite{PhysRevD.27.233,kitazawa2002chiral,bohr2009color,tsue2013interplay}
\begin{align}
\label{eq:1}
    H=H_{B}+H_{Q}+H_{T},
\end{align}
where $H_{B(Q)}$ and $H_T$
are the baryon (quark) Hamiltonian and the tunneling Hamiltonian, respectively.
The baryon Hamiltonian reads
\begin{align}
\label{eq:2}
    H_{B}=&
    \sum_{{\rm B}}
    \left(\xi_{\bm{K},{\rm B}}
    B_{\bm{K}\sigma_B\tau_B}^\dag
    B_{\bm{K}\sigma_B\tau_B}
    -\xi_{\bm{K},{\rm \bar{B}}}
    \bar{B}_{\bm{K}\sigma_B\tau_B}^\dag
    \bar{B}_{\bm{K}\sigma_B\tau_B}
    \right)\cr
    &+V_{BB}+V_{B\bar{B}}+V_{ \bar{B}\bar{B}},
\end{align}
where $\xi_{\bm{K},{ B}}=\sqrt{K^2+M_{ B}^2}-\mu_{B}$
and $\xi_{\bm{K},{ \bar{B}}}=\sqrt{K^2+M_{ B}^2}+\mu_{B}$
are the kinetic energies of a baryon and an anti-baryon with momentum $\bm{K}$, mass $M_{B}$ and chemical potential $\mu_{B}$. 
$B_{\bm{K}\sigma_B\tau_B}^{(\dag)}$ and
$\bar{B}_{\bm{K}\sigma_B\tau_B}^{(\dag)}$ are
the annihilation (creation) operators of a baryon and an anti-baryon with spin $\sigma_B$ and isospin $\tau_B$.
For convenience, we introduced the summation of baryon quantum numbers as
\begin{align}
   \sum_{{\rm B}}=\sum_{\bm{K}}\sum_{\sigma_B}\sum_{\tau_B}. 
\end{align}
$V_{BB}$, $V_{B\bar{B}}$, and $V_{ \bar{B}\bar{B}}$ in Eq.~\eqref{eq:2} represent the baryon--baryon, baryon--anti-baryon, and anti-baryon--anti-baryon interactions, respectively.
In a similar way, we introduce the quark Hamiltonian
\begin{align}
\label{eq:2.4}
    H_{Q}=&
    \sum_{\rm Q}
    \left(\xi_{\bm{k},{Q}}
    q_{\bm{k}\sigma\tau a}^\dag
    q_{\bm{k}\sigma\tau a}
    -\xi_{\bm{k},{\rm \bar{Q}}}
    \bar{q}_{\bm{k}\sigma\tau a}^\dag
    \bar{q}_{\bm{k}\sigma\tau a}
    \right)\cr
    &+V_{QQ}+V_{Q\bar{Q}}+V_{ \bar{Q}\bar{Q}},
\end{align}
where $\xi_{\bm{k},{ Q}}=\sqrt{k^2+M_{ Q}^2}-\mu_{ Q}$ and $\xi_{\bm{k},\bar{ Q}}=\sqrt{k^2+M_{ Q}^2}+\mu_{Q}$ are the kinetic energy of a quark and an anti-quark with momentum $\bm{k}$, mass $M_{Q}$, and chemical potential $\mu_{ Q}$ respectively.
The annihilation (creation) operators $q_{\bm{k}\sigma\tau a}^{(\dag)}$ and $\bar{q}_{\bm{k}\sigma\tau a}^{(\dag)}$ of a quark and an anti-quark involve the spin $\sigma$, flavor (or isospin) $\tau$, and color $a$.
As in the case of baryons, we have introduced the short-nand notation of the summation of quark quantum numbers as
\begin{align}
    \sum_{\rm Q}=\sum_{\bm{k}}\sum_{\sigma}\sum_{\tau}\sum_{a}
\end{align}
$V_{QQ}$, $V_{Q\bar{Q}}$ and $V_{ \bar{Q}\bar{Q}}$ are the quark--quark, quark--anti-quark, and anti-quark--anti-quark interactions, respectively.
Then, we consider the tunneling Hamiltonian at the hadron-quark interface given by
\begin{align}
\label{eq:2.5}
    H_{T}
    &=
    \sum_{{\rm B},{\rm Q}_1,{\rm Q}_2,{\rm Q}_3}
    \mathcal{T}_{(\bm{K},\sigma_B,\tau_B,\{\bm{k}_i,\sigma_i,\tau_i,a_i\})}
    B_{\bm{K}\sigma_B\tau_B}^\dag \cr
&
\quad
\times    \varepsilon_{a_1a_2a_3}
    q_{\bm{k}_1\sigma_1\tau_1a_1}
    q_{\bm{k}_2\sigma_2\tau_2}
    q_{\bm{k}_3\sigma_3\tau_3}
    +{\rm h.c.}\cr
    &\,+
    \sum_{{\rm B},{\rm Q}_1,{\rm Q}_2,{\rm Q}_3}
    \bar{\mathcal{T}}_{(\bm{K},\sigma_B,\tau_B,\{\bm{k}_i,\sigma_i,\tau_i,a_i\})}
    \bar{B}_{\bm{K}\sigma_B\tau_B}^\dag \cr
&
\quad
\times    \varepsilon_{a_1a_2a_3}
    \bar{q}_{\bm{k}_1\sigma_1\tau_1a_1}
    \bar{q}_{\bm{k}_2\sigma_2\tau_2}
    \bar{q}_{\bm{k}_3\sigma_3\tau_3}
    +{\rm h.c.}\cr
\end{align}
where
$\varepsilon_{a_1a_2a_3}$ is the anti-symmetric tensor for color degrees of freedom and
we introduced 
\begin{align}
\sum_{{\rm Q}_1,{\rm Q}_2,{\rm Q}_3}=
    \sum_{\bm{k}_1,\bm{k}_2,\bm{k}_3}
    \sum_{\sigma_1,\sigma_2,\sigma_3}
    \sum_{\tau_1,\tau_2,\tau_3}
    \sum_{a_1,a_2,a_3}.
\end{align}
$\mathcal{T}_{(\bm{K},\sigma_B,\tau_B,\{\bm{k}_i,\sigma_i,\tau_i,a_i\})}$ and  $\bar{\mathcal{T}}_{(\bm{K},\sigma_B,\tau_B,\{\bm{k}_i,\sigma_i,\tau_i,a_i\})}$
are the tunneling amplitudes of a color-singlet baryon to three quarks
and that of a color-singlet anti-baryon to three anti-quarks across the interface.
In addition to Eq.~\eqref{eq:2.5},
the tunneling Hamiltonian for the conversion between a meson and a quark--anti-quark pair~\cite{juzaki2020andreev} can be important in the low-density regime.
In this paper, we focus on the process given by Eq.~\eqref{eq:2.5} assuming the high-density regime where the quark-anti-quark pair creation is strongly suppressed.

We note that the tunneling amplitude can be estimated from the three-quark confinement force, which can be found in the lattice QCD simulation~\cite{PhysRevLett.86.18,PhysRevD.65.114509}, 
as well as
the overlap integral of the baryon and three-quark wave functions at the tunneling region in analogy with strongly-interacting condensed matter systems~\cite{PhysRevA.106.033310,PhysRevB.108.155303}.
We do not go into the precise estimation of the tunneling amplitudes because this paper is dedicated to the derivation of the tunneling transport in a neutron star.

Incidentally, while we do not present explicit forms of the interaction terms,
we consider the BCS pairing states in baryon and quark matter with constant pairing gaps in the numerical demonstration.
These assumptions correspond to the case with contact-type $s$-wave baryon-baryon and quark-quark attractive interactions.
The accurate expressions of the interactions (e.g., realistic nuclear forces~\cite{PhysRevC.48.792,PhysRevC.51.38,PhysRevC.63.024001}) may help us estimate the momentum-dependent nucleon pairing gaps quantitatively.
However, the contact-type-interaction model is sufficient for our purpose.

\section{Schwinger-Keldysh approach to the tunneling transport across the interface}
\label{sec:formalism}

In the following, we employ the so-called relativized quark model~\cite{PhysRevD.34.2809} where the excitations of antiparticles (i.e., anti-baryons and anti-quarks) and mesons are ignored.
This approximation is valid at high densities due to the large chemical potential, or at the heavy quark-mass limit where the non-relativistic treatment is justified.
Nevertheless, it should be noted that the relativized quark model can successfully describe the observed baryon masses even under this approximation~\cite{PhysRevD.34.2809}. 

First, we introduce the tunneling current and friction operators acting on the interface. 
Based on the Heisenberg equation,
one can introduce the tunneling current operator
\begin{align}
    \hat{I}&=-\frac{d}{dt}\rho_{B}=i\left[\rho_{ B},H\right],
\end{align}
where
\begin{align}
    \rho_{ B}=
    \sum_{\rm B}
    B_{\bm{K}\sigma_B\tau_B}^\dag B_{\bm{K}\sigma_B\tau_B}.
\end{align}
is the baryon number-density operator.
$\hat{I}$ represents the changing rate of $\rho_{\rm B}$ due to the tunneling of baryons toward dense quark matter side via $H_{T}$.
Rewriting $H_{T}$ as
\begin{align}
    H_{T} =\hat{\Gamma}_I+\hat{\Gamma}_I^\dagger,
\end{align}
with
\begin{align}
    \hat{\Gamma}_I&=
    \sum_{{B}, {\rm Q}_1,{\rm Q}_2,{\rm Q}_3}
    \mathcal{T}_{(\bm{K},\sigma_B,\tau_B,\{\bm{k}_i,\sigma_i,\tau_i,a_i\})}
    B_{\bm{K}\sigma_B\tau_B}^\dag \cr
    &
\quad\quad
\times    \varepsilon_{a_1a_2a_3}
    q_{\bm{k}_1\sigma_1\tau_1a_1}
    q_{\bm{k}_2\sigma_2\tau_2}
    q_{\bm{k}_3\sigma_3\tau_3},
\end{align}
we obtain
\begin{align}
    \hat{I}=i(\hat{\Gamma}_I-\hat{\Gamma}_I^\dagger).
\end{align}

In a similar manner, one may introduce the friction operator as
\begin{align}
    \hat{\bm{F}}=-\frac{d\bm{P}_{B}}{dt}
    =i[\bm{P}_{B},H],
\end{align}
where 
\begin{align}
    \bm{P}_{B}
    &=
    \sum_{\rm B}
    \bm{K}B_{\bm{K}\sigma_B\tau_B}^\dag B_{\bm{K}\sigma_B\tau_B}
\end{align}
is the total momentum operator of baryon matter in the rest frame of quark matter.
In this regard, $\bm{P}_{B}$ can also be regarded as the relative momentum between baryon and quark bulk systems.
Using the anti-commutation relation of the baryon operators, we obtain
\begin{align}
    \hat{\bm{F}}=i(\hat{\bm{\Gamma}}_F-\hat{\bm{\Gamma}}_F^\dagger),
\end{align}
where
\begin{align}
    \hat{\bm{\Gamma}}_F&=
    \sum_{{\rm B},{\rm Q}_1,{\rm Q}_2,{\rm Q}_3}
    \mathcal{T}_{(\bm{K},\sigma_B,\tau_B,\{\bm{k}_i,\sigma_i,\tau_i,a_i\})}\bm{K}
B_{\bm{K}\sigma_B\tau_B}^\dag \cr
    &
\quad\quad
\times    \varepsilon_{a_1a_2a_3}
    q_{\bm{k}_1\sigma_1\tau_1a_1}
    q_{\bm{k}_2\sigma_2\tau_2}
    q_{\bm{k}_3\sigma_3\tau_3}.
\end{align}

\begin{figure}[t]
    \centering
    \includegraphics[width=\linewidth]{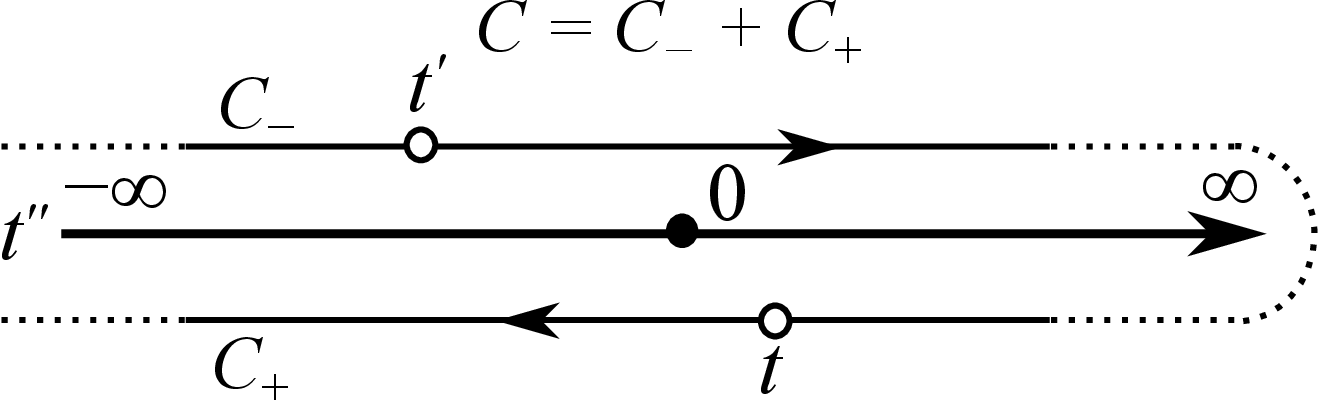}
    \caption{Keldysh contour $C=C_-+C_+$ along the $t''$ axis in Eq.~\eqref{eq:3.14}, consisting of a forward branch $C_-$ and a backward one $C_{+}$.
    Two time parameters $t$ and $t'$ in $\langle\hat{I}(t,t')\rangle$ and $\langle\hat{\bm{F}}(t,t')\rangle$ locate on $C_+$ and $C_-$, respectively. }
    \label{fig:contour}
\end{figure}

Under the non-equilibrium condition,
we are interested in the two-time expectation values of the tunneling current and the friction given by
\begin{widetext}
\begin{align}
\label{eq:I}
    \langle \hat{I}(t,t')\rangle
    &=i\langle\hat{\Gamma}_I(t,t')\rangle +{\rm h.c.}\cr
    &=i
    \sum_{{\rm B},{\rm Q}_1,{\rm Q}_2,{\rm Q}_3}
    \mathcal{T}_{(\bm{K},\sigma_B,\tau_B,\{\bm{k}_i,\sigma_i,\tau_i,a_i\})}
    \varepsilon_{a_1a_2a_3}\cr
    &
    \quad
    \times
     \left\langle
    T_C\left[\hat{S}_C
    B_{\bm{K}\sigma_B\tau_B}^{\dag (H)}(t)
    q_{\bm{k}_1\sigma_1\tau_1a_1}^{(H)}(t')
    q_{\bm{k}_2\sigma_2\tau_2a_2}^{(H)}(t')
    q_{\bm{k}_3\sigma_3\tau_3a_3}^{(H)}(t')\right]\right\rangle +{\rm h.c.},
\end{align}
\begin{align}
    \langle \hat{\bm{F}}(t,t')\rangle
    &=i\langle\hat{\bm{\Gamma}}_F(t,t')\rangle +{\rm h.c.}\cr
    &=i
    \sum_{{\rm B},{\rm Q}_1,{\rm Q}_2,{\rm Q}_3}
    \mathcal{T}_{(\bm{K},\sigma_B,\tau_B,\{\bm{k}_i,\sigma_i,\tau_i,a_i\})}\bm{K}
    \varepsilon_{a_1a_2a_3}\cr
    &
    \quad
    \times
     \left\langle
    T_C\left[\hat{S}_C
    B_{\bm{K}\sigma_B\tau_B}^{\dag (H)}(t)
    q_{\bm{k}_1\sigma_1\tau_1a_1}^{(H)}(t')
    q_{\bm{k}_2\sigma_2\tau_2a_2}^{(H)}(t')
    q_{\bm{k}_3\sigma_3\tau_3a_3}^{(H)}(t')\right]\right\rangle +{\rm h.c.},
\end{align}
\end{widetext}
with the contour ordering product $T_C$~\cite{Keldysh},
where 
\begin{align}
B_{\bm{K}\sigma_B\tau_B}^{\dag (H)}(t)=e^{iH_{B}t}B_{\bm{K}\sigma_B\tau_B}^{\dag} e^{-iH_{B}t}    
\end{align}
 and 
\begin{align}
q_{\bm{k}_j\sigma_j\tau_ja_j}^{(H)}(t)=e^{iH_{Q}t}q_{\bm{k}_j\sigma_j\tau_ja_j}e^{-iH_Qt}    
\end{align} 
 are the Heisenberg representations of the operators and
\begin{align}
\label{eq:3.14}
    \hat{S}_C=\exp\left[-i\int_Cdt'' H_{T}^{(H)}(t'')\right]
\end{align}
is the $S$-matrix operator with respect to the tunneling Hamiltonian $H_{ T}^{(H)}(t)=e^{i(H_B+H_Q)t}H_T e^{-i(H_B+H_Q)t}$ in the interaction representation.
The time integration contour $C$ in Eq.~\eqref{eq:3.14} is called the Keldysh contour~\cite{Keldysh} that runs from $t''=-\infty$ to $t''=\infty$ (forward branch $C_{-}$) and then returns back to $t''=-\infty$ (backward branch $C_{+}$) as shown in Fig.~\ref{fig:contour}.
One can perform the perturbative expansion of $\langle \hat{I}(t,t')\rangle$ and
$\langle \hat{\bm{F}}(t,t')\rangle$ with respect to $\mathcal{T}$ by expanding $\hat{S}_C$.
In particular, the Josephson current arises at the second order of $H_{T}$ at the superfluid-superfluid interface~\cite{PhysRevB.54.7366}.
Moreover, the Andreev reflection process can occur at the fourth order of $H_{ T}$ at the normal-superfluid interface~\cite{PhysRevB.54.7366,zhang2023dominant}.

The tunneling current can be induced by the chemical potential differences between baryon matter and quark matter, denoted as $\mu_{B}-3\mu_{Q}$ (noting that $\mu_{B}\neq 3\mu_{Q}$ in the presence of the chemical potential bias at the interface).
In this regard, we express the averaged value in terms of thermal Green's function where the operators are written in the grand-canonical Heisenberg representation with
\begin{align}
K_{B}=H_{B}-\mu_{B}\rho_B-\bm{v}_{ B}\cdot\bm{P}_{B},    
\end{align}
and
\begin{align}
    K_{Q}=H_{Q}-\mu_{Q}\rho_{Q},
\end{align}
where $\rho_Q$ is the quark number density.
Moreover, we have introduced the velocity $\bm{v}_{B}$ of baryon matter in the rest frame of quark matter, which can be regarded as the Lagrange multiplier for a given $\bm{P}_{B}$. 
The velocity difference between two bulk systems can also induce the tunneling transport even without the chemical potential bias (i.e., $\mu_{ B}=3\mu_{Q}$) as discussed in the spin system~\cite{PhysRevB.105.L020302}.
As shown in Fig.~\ref{fig:interface},
the differential rotation of a neutron star can be introduced as the local linear motion with $\bm{v}_{B}$ parallel to the interface.
The annihilation operators in the Heisenberg representation of $K_{\rm B/Q}$ are given by
\begin{align}
    B_{\bm{K}\sigma_B\tau_B}^{(K)}(t)&=e^{iK_Bt}B_{\bm{K}\sigma_B\tau_B}e^{-iK_Bt}\cr
    &=e^{i(\mu_{B}+\bm{v}_{ B}\cdot\bm{K})t} B_{\bm{K}\sigma_B\tau_B}^{(H)}(t),
\end{align}
\begin{align}
    q_{\bm{k}_j\sigma_j\tau_j a_j}^{(K)}(t)&=e^{iK_Qt}q_{\bm{k}_j\sigma_j\tau_j a_j}e^{-iK_Qt}\cr
    &=
    e^{i\mu_{ Q}t}q_{\bm{k}_j\sigma_j\tau_j a_j}^{(H)}(t).
\end{align}
Using them, we obtain the perturbative expression of the tunneling current and the friction in terms of the grand-canonical Heisenberg representations as
\begin{widetext}
\begin{align}
\label{eq:current}
    \langle\hat{I}(t,t')\rangle&=
    i\sum_{{\rm B},{\rm Q}_1,{\rm Q}_2,{\rm Q}_3}
    \mathcal{T}_{(\bm{K},\sigma_B,\tau_B,\{\bm{k}_i,\sigma_i,\tau_i,a_i\})}
    \varepsilon_{a_1a_2a_3}
    e^{i(\mu_{\rm B}+\bm{v}_{\rm B}\cdot\bm{K}) t}
    e^{-3i\mu_{\rm Q} t'}
    \cr
    &\quad\quad\times
    \left\langle
    T_C\left[\hat{S}_C
    B_{\bm{K}\sigma_B\tau_B}^{\dag (K)}(t)
    q_{\bm{k}_1\sigma_1\tau_1a_1}^{(K)}(t')
    q_{\bm{k}_2\sigma_2\tau_2}^{(K)}(t')
    q_{\bm{k}_3\sigma_3\tau_3}^{(K)}(t')\right]\right\rangle_0 +{\rm h.c.},
\end{align}
\begin{align}
\label{eq:friction}
    \langle \hat{\bm{F}}(t,t')\rangle
    &=i
    \sum_{{\rm B},{\rm Q}_1,{\rm Q}_2,{\rm Q}_3}
    \mathcal{T}_{(\bm{K},\sigma_B,\tau_B,\{\bm{k}_i,\sigma_i,\tau_i,a_i\})}\bm{K}
    \varepsilon_{a_1a_2a_3}
    e^{i(\mu_{\rm B}+\bm{v}_B\cdot\bm{K})t}
    e^{-3i\mu_{\rm Q}t'}
    \cr
    &
    \quad
    \times
     \left\langle
    T_C\left[\hat{S}_C
    B_{\bm{K}\sigma_B\tau_B}^{\dag (K)}(t)
    q_{\bm{k}_1\sigma_1\tau_1a_1}^{(K)}(t')
    q_{\bm{k}_2\sigma_2\tau_2a_2}^{(K)}(t')
    q_{\bm{k}_3\sigma_3\tau_3a_3}^{(K)}(t')\right]\right\rangle +{\rm h.c.}.
\end{align}
\end{widetext}
In this way, one can discuss the tunneling current, that has been well studied in condensed-matter systems, and the friction relevant to the differential rotation in a neutron star.

\section{Josephson tunneling current at the hadron-quark interface}
\begin{figure}[t]
    \centering
    \includegraphics[width=0.9\linewidth]{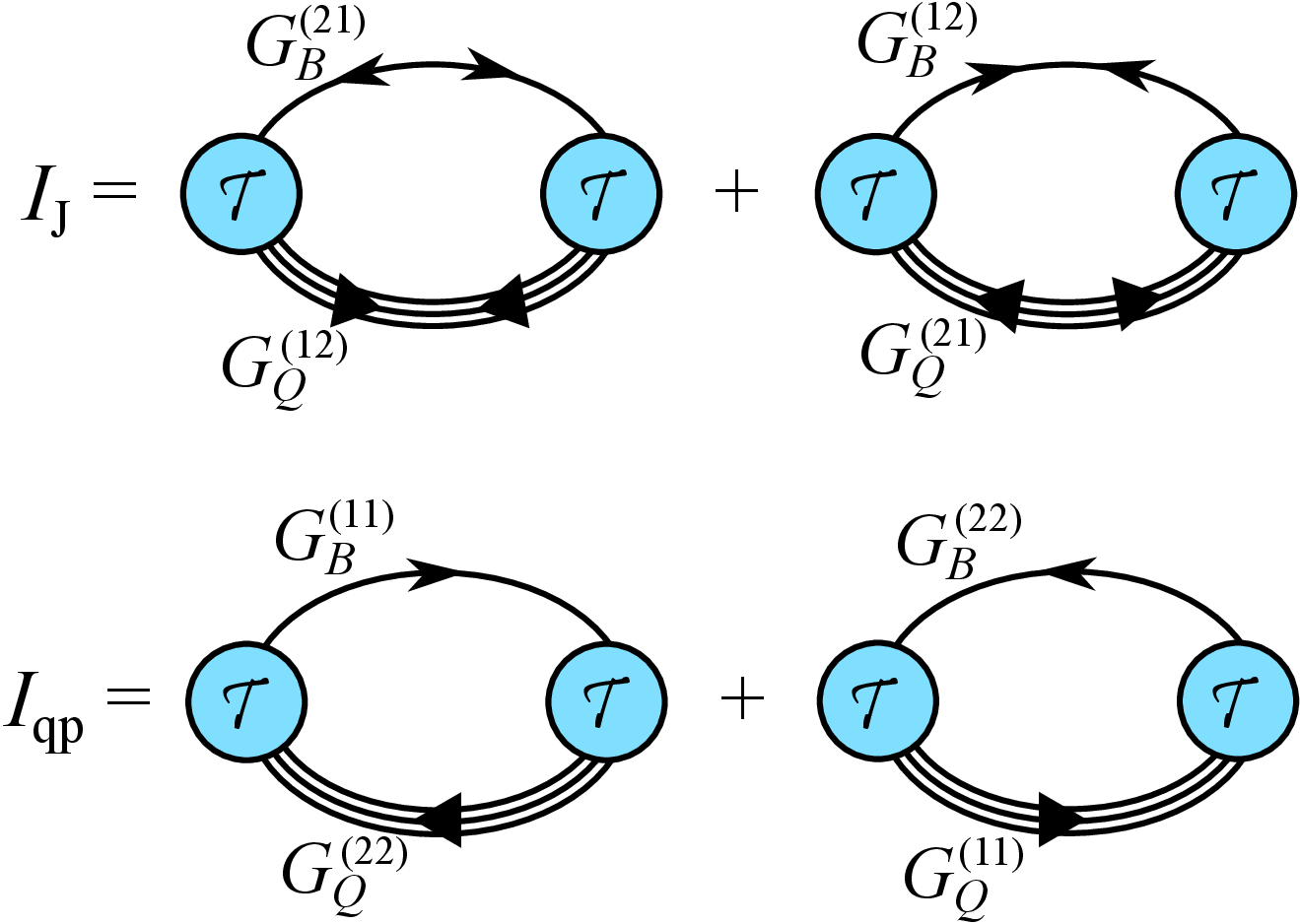}
    \caption{Feynman diagrams for the Josephson tunneling current $I_{\rm J}$ and the quasiparticle tunneling current $I_{\rm qp}$. 
    The single- and triple-solid lines show baryon and three-quark propagators, respectively, within the Nambu-Gor'kov representation.
    The circle represents the tunneling coupling $\mathcal{T}$.}
    \label{fig:diagram}
\end{figure}

For the sake of demonstration of our field theoretical framework, we evaluate the Josephson tunneling current within the second order perturbation of $H_{T}$ in Eq.~\eqref{eq:current}.
For convenience,
we introduce 
\begin{align}
    Q_{\{\bm{k}_i,\sigma_i,\tau_i,a_i\}}^{(K)}(t')
    =&\varepsilon_{a_1a_2a_3}
    q_{\bm{k}_1\sigma_1\tau_1a_1}^{(K)}(t')
    q_{\bm{k}_2\sigma_2\tau_2a_2}^{(K)}(t')\cr
    &\times
    q_{\bm{k}_3\sigma_3\tau_3a_3}^{(K)}(t').
\end{align}
Then, we obtain the leading-order contribution of the tunneling current given by
\begin{widetext}
\begin{align}
    \langle\hat{I}_1(t,t')\rangle&=
    \sum_{{\rm B},{\rm B}'}
    \sum_{{\rm Q}_1,{\rm Q}_2,{\rm Q}_3,{\rm Q}_1',{\rm Q}_2',{\rm Q}_3'}
    \int_C dt_1
    \mathcal{T}_{(\bm{K},\sigma_B,\tau_B,\{\bm{k}_i,\sigma_i,\tau_i,a_i\})}
    \mathcal{T}_{(\bm{K}'\sigma_B',\tau_B',\{\bm{k}_i',\sigma_i',\tau_i,a_i'\})}
    e^{i(\mu_{B} +\bm{v}_{ B}\cdot\bm{K})t}
    e^{-3i\mu_{ Q} t'}
    \cr
    &\quad\quad\times e^{i(\mu_{ B}+\bm{v}_{ B}\cdot\bm{K}-3\mu_{ Q})t_1}
    \left\langle
    T_C\left[
    B_{\bm{K}\sigma_B\tau_B}^{\dag (K)}(t)
    Q_{\{\bm{k}_i,\sigma_i,\tau_i,a_i\}}^{(K)}(t')
    B_{\bm{K}'\sigma_B'\tau_B'}^{\dag (K)}(t_1)
    Q_{\{\bm{k}_i',\sigma_i',\tau_i',a_i'\}}^{(K)}(t_1)
    \right]\right\rangle_0 +{\rm h.c.}\cr
    &+
     \sum_{{\rm B},{\rm B}'}
    \sum_{{\rm Q}_1,{\rm Q}_2,{\rm Q}_3,{\rm Q}_1',{\rm Q}_2',{\rm Q}_3'}
    \int_C dt_1
    \mathcal{T}_{(\bm{K},\sigma_B,\tau_B,\{\bm{k}_i,\sigma_i,\tau_i,a_i\})}
    \mathcal{T}_{(\bm{K}',\sigma_B',\tau_B'\{\bm{k}_i',\sigma_i',\tau_i,a_i'\})}^*
    e^{i(\mu_{B}+\bm{v}_{ B}\cdot\bm{K}) t}
    e^{-3i\mu_{ Q} t'}
    \cr
    &\quad\quad\times e^{-i(\mu_{ B}+\bm{v}_{ B}\cdot\bm{K}-3\mu_{ Q})t_1}
    \left\langle
    T_C\left[
    B_{\bm{K}\sigma_B\tau_B}^{\dag (K)}(t)
    Q_{\{\bm{k}_i,\sigma_i,\tau_i,a_i\}}^{(K)}(t')
    Q_{\{\bm{k}_i',\sigma_i',\tau_i',a_i'\}}^{\dag(K)}(t_1)
    B_{\bm{K}'\sigma_B'\tau_B'}^{(K)}(t_1)
    \right]\right\rangle_0 +{\rm h.c.}\cr
    &\equiv I_{\rm J}(t,t')+I_{\rm qp}(t,t'),
\end{align}
\end{widetext}
where one may find two terms corresponding to the Josephson tunneling current $I_{\rm J}(t,t')$ and the quasiparticle tunneling current $I_{\rm qp}(t,t')$.
While $I_{\rm J}(t,t')$ involves the anomalous Green's functions (i.e., $\langle T_CQ^{(K)}(t')Q^{(K)}(t_1)\rangle $ and $\langle T_C B^{\dag (K)}(t_1) B^{\dag(K)}(t)\rangle$) specific to the superfluid phases~\cite{PhysRevLett.10.486},
$I_{\rm qp}(t,t')$ is written in terms of only normal Green's functions
as shown in Fig.~\ref{fig:diagram}
and thus becomes nonzero even in the normal phase.

Here we focus on $I_{\rm J}(t,t')$ which can be nonzero even without the biases of the chemical potential and the velocity if the phases of two superfluid order parameters are different.
Using the Wick theorem, we get
\begin{widetext}
\begin{align}\label{IJ1}
    I_{\rm J}(t,t')
    =&
     \sum_{{\rm B},{\rm B}'}
    \sum_{{\rm Q}_1,{\rm Q}_2,{\rm Q}_3,{\rm Q}_1',{\rm Q}_2',{\rm Q}_3'}
    \int_C dt_1
    \mathcal{T}_{(\bm{K},\sigma_B,\tau_B,\{\bm{k}_i,\sigma_i,\tau_i,a_i\})}
    \mathcal{T}_{(\bm{K}',\sigma_B',\tau_B',\{\bm{k}_i',\sigma_i',\tau_i,a_i'\})}
    e^{i(\mu_B +\bm{v}_{\rm B}\cdot\bm{K})t}
    e^{-3i\mu_Q t'}
    \nonumber\\
    &\times e^{i(\mu_B+\bm{v}_{ B}\cdot\bm{K}-3\mu_Q)t_1}
    \left\langle
    T_C\,
    Q_{\{\bm{k}_i,\sigma_i,\tau_i,a_i\}}^{(K)}(t')Q_{\{\bm{k}_i',\sigma_i',\tau_i',a_i'\}}^{(K)}(t_1)\right\rangle_0
    \left\langle T_C\, B_{\bm{K}'\sigma_B'\tau_B'}^{\dag (K)}(t_1)
    B_{\bm{K}\sigma_B\tau_B}^{\dag (K)}(t)\right\rangle_0 +{\rm h.c.}
\end{align}
\end{widetext}
Furthermore, we introduce the contour-time-ordered Green's function of quarks and baryons with the $2\times 2$ Nambu-Gor'kov representation
\begin{align}
    &iG_{Q,\{\bm{k}_i,\sigma_i,\tau_i,a_i\}}(t,t')\cr
    &=\left\langle T_C \left[A_{Q,\{\bm{k}_i,\sigma_i,\tau_i,a_i\}}(t)A^\dagger_{Q,\{\bm{k}_i,\sigma_i,\tau_i,a_i\}}(t')\right]\right\rangle_0,
\end{align}
\begin{align}
    &iG_{B,\bm{K},\sigma_B,\tau_B}(t,t')
    \cr
    &=\left\langle T_C \left[A_{B,\bm{K},\sigma_B,\tau_B}(t)A^\dagger_{B,\bm{K},\sigma_B,\tau_B}(t')\right]\right\rangle_0,
\end{align}
where $A_{Q,\{\bm{k}_i,\sigma_i,\tau_i,a_i\}}(t)=\left(Q_{\{\bm{k}_i,\sigma_i,\tau_i,a_i\}}(t),Q_{\{-\bm{k}_i,\bar{\sigma}_i,\bar{\tau}_i,\bar{a}_i\}}^\dagger(t)\right)^T$, and $A_{B,\bm{K},\sigma_B,\tau_B}(t)=\left(B_{\bm{K},\sigma_B,\tau_B}(t),B_{-\bm{K},\bar{\sigma}_B,\bar{\tau}_B}^\dagger(t)\right)^T$ (the labels with bar implies the quantum states of pairing partner. For example, for spin-singlet neutron pairing, we take  $\bar{\sigma}_{\rm B}=-\sigma_{\rm B}$, $\tau_{\rm B}=\bar{\tau}_{\rm B}$.).
For simplicity, we assume that the tunneling amplitude does not depend on quantum numbers as $\mathcal{T}_{\bm{K},\sigma_B,\tau_B,\{\bm{k}_i,\sigma_i,\tau_i,a_i\}}\simeq \mathcal{T}$ and moreover $\mathcal{T}$ is taken to be a real value.
In terms of contour-time-ordered Green's functions, one can write
    \begin{align}\label{IJ2}
    I_{\rm J}(t,t')
    &=-
     \sum_{{\rm B}}
    \sum_{{\rm Q}_1,{\rm Q}_2,{\rm Q}_3}
    \int_C dt_1\,
    \mathcal{T}^2
    \cr
    &
    \quad
    \times
    e^{i(\mu_B+\bm{v}_{\rm B}\cdot\bm{K}) t} e^{-3i\mu_Q t'} e^{i(\mu_B+\bm{v}_{\rm B}\cdot\bm{K}-3\mu_Q)t_1} \cr
    &
    \quad
    \times G^{(12)}_{Q,\{\bm{k}_i,\sigma_i,\tau_i,a_i\}}(t',t_1)G^{(21)}_{B,\bm{K},\sigma_B,\tau_B}(t_1,t)\cr
    &\quad\quad+{\rm h.c.},
\end{align}
where $G^{(12)}_Q$ and $G^{(21)}_B$ are off-diagonal components of their Green's functions.  We note that $t$ and $t'$ are time parameters respectively locating on $C_+$ and $C_-$ in Fig.~\ref{fig:contour}.
Denoting the time parameters locating on $C_+$ and $C_-$ respectively by $t^+$ and $t^-$, we introduce the greater and lesser Green's functions as 
\begin{align}
    &iG^>_{Q,\{\bm{k}_i,\sigma_i,\tau_i,a_i\}}(t^+_1,t^-_2)
    \cr
    &=\left\langle A_{Q,\{\bm{k}_i,\sigma_i,\tau_i,a_i\}}(t^+_1)A^\dagger_{Q,\{\bm{k}_i,\sigma_i,\tau_i,a_i\}}(t^-_2)\right\rangle_0,\\
    &iG^>_{B,\bm{K},\sigma_B,\tau_B}(t^+_1,t^-_2)
    \cr
    &=\left\langle A_{B,\bm{K},\sigma_B,\tau_B}(t^+_1)A^\dagger_{B,\bm{K},\sigma_B,\tau_B}(t^-_2)\right\rangle_0,\\
    &iG^<_{Q,\{\bm{k}_i,\sigma_i,\tau_i,a_i\}}(t^-_1,t^+_2)\cr
    &=\left\langle A_{Q,\{\bm{k}_i,\sigma_i,\tau_i,a_i\}}(t_1^-)A^\dagger_{Q,\{\bm{k}_i,\sigma_i,\tau_i,a_i\}}(t_2^+)\right\rangle_0,\\
    &iG^<_{B,\bm{K},\sigma_B,\tau_B}(t^-_1,t^+_2)\cr
    &=\left\langle A_{B,\bm{K},\sigma_B,\tau_B}(t^-_1)A^\dagger_{B,\bm{K},\sigma_B,\tau_B}(t^+_2)\right\rangle_0.
\end{align}
Also, assuming steady-state transport where the time scale of the bulk dynamics is sufficiently slow compared to the tunneling dynamics (that occurs during the relative time $t-t'$), we consider the limit of $t'\rightarrow t$.
Then, the Josepshon current is written as a single-time function as
\begin{widetext}
\begin{align}\label{IJ3}
    I_{\rm J}(t)
    =&-2
    \sum_{\rm B}
    \sum_{\rm Q_1,Q_2,Q_3}
    \int_{-\infty}^{\infty} dt_1
    \operatorname{Re}\Big\{\mathcal{T}^2
    e^{i[\Delta\mu (t+t_1)+\Delta\phi]}\cr 
    &\Big[G^{(12){\rm ret.}}_{Q,\{\bm{k}_i,\sigma_i,\tau_i,a_i\}}(t,t_1)G^{(21)<}_{B,\bm{K},\sigma_B,\tau_B}(t_1,t)+G^{(12)<}_{Q,\{\bm{k}_i,\sigma_i,\tau_i,a_i\}}(t,t_1)G^{(21){\rm adv.}}_{B,\bm{K},\sigma_B,\tau_B}(t_1,t)\Big]\Big\},
\end{align}
\end{widetext}
where $\Delta\mu=\mu_B+\bm{v}_{ B}\cdot\bm{K}-3\mu_Q$ is the external bias parameter. $\Delta\phi=\phi_B-3
\phi_Q$
is the phase difference of the gap parameters between the hadron and quark sides, where the gap parameters are respectively expressed as $\Delta_B=|\Delta_B|e^{-i\phi_B}$ and $\Delta_Q=|\Delta_Q|e^{-i\phi_Q}$.

Using the relation between fermionic lesser (greater) Green's functions and retarded Green's functions in thermal-equilibrium bulk systems given by~\cite{rammer2011quantum} 
\begin{align}
G^<(\omega)&=-2i\operatorname{Im}\big[G^{\rm ret.}(\omega)\big]f(\omega), 
\end{align}
\begin{align}
    G^>(\omega)&=2i\operatorname{Im}\big[G^{\rm ret.}(\omega)\big][1-f(\omega)],
\end{align}
with the Fermi distribution function $f(\omega)=1/(e^{\omega/T}+1)$, 
we can rewrite $I_{\rm J}(t)$
 as 
\begin{widetext}
\begin{align}\label{IJ12}
    I_{\rm J}(t)
    =4\sum_{\bm{K},\sigma_B,\tau_B}
    \sum_{\{\bm{k}_i,\sigma_i,\tau_i,a_i\}}
    &\int_{-\infty}^{\infty} \frac{d\omega}{2\pi}
    \Big\{\mathcal{T}^2
    \operatorname{Im}G^{(12){\rm ret.}}_{Q,\{\bm{k}_i,\sigma_i,\tau_i,a_i\}}(\omega-\Delta\mu)\operatorname{Im}G^{(21){\rm ret.}}_{B,\bm{K},\sigma_B,\tau_B}(\omega)\nonumber\\
    &\times[
    f(\omega-\Delta\mu)-f(\omega)]\cos(2\Delta\mu t+\Delta\phi)\nonumber\\
    &-\mathcal{T}^2
    \Big[\operatorname{Re}G^{(12){\rm ret.}}_{Q,\{\bm{k}_i,\sigma_i,\tau_i,a_i\}}(\omega-\Delta\mu)\operatorname{Im}G^{(21){\rm ret.}}_{B,\bm{K},\sigma_B,\tau_B}(\omega)f(\omega)\nonumber\\
    &+\operatorname{Im}G^{(12){\rm ret.}}_{Q,\{\bm{k}_i,\sigma_i,\tau_i,a_i\}}(\omega-\Delta\mu)\operatorname{Re}G^{(21){\rm ret.}}_{B,\bm{K},\sigma_B,\tau_B}(\omega)f(\omega-\Delta\mu)\Big]\sin(2\Delta\mu t+\Delta\phi)\Big\}.
\end{align}
\end{widetext}
When one takes $\Delta\mu=0$, the first term of 
Eq.~(\ref{IJ12})
vanishes, while the second term remains as a DC current proportional to $\sin{(\Delta\phi)}$.

As we stated in the beginning of this section,
we consider the BCS pairing states of particle states at high densities without anti-particle contributions.
We use the retarded Nambu-Gor'kov Green's function of a baryon and a quark as
\begin{align}
    G^{(11){\rm ret.}}_{B,\bm{K},\sigma_B,\tau_B}(\omega)=\frac{u_{\bm{K},B}^2}{\omega+i\eta -E_{\bm{K},B}}+\frac{v_{\bm{K},B}^2}{\omega+i\eta +E_{\bm{K},B}},
\end{align}
\begin{align}
    g^{(11){\rm ret.}}_{Q,\bm{k},\sigma,\tau,a}(\omega)=\frac{u_{\bm{k},Q}^2}{\omega+i\eta -E_{\bm{k},Q}}+\frac{v_{\bm{k},Q}^2}{\omega+i\eta +E_{\bm{k},Q}},
\end{align}
\begin{align}
    &G^{(12){\rm ret.}}_{B,\bm{K},\sigma_B,\tau_B,a_B}(\omega)=-\frac{|\Delta_B|}{2E_{\bm{K},B}}\cr
    &\times \bigg[\frac{1}{\omega+i\eta-E_{\bm{K},B}}-\frac{1}{\omega+i\eta+E_{\bm{K},B}}\bigg],
\end{align}
\begin{align}
    &g^{(12){\rm ret.}}_{Q,\bm{k},\sigma,\tau,a}(\omega)=-\frac{|\Delta_Q|}{2E_{\bm{k},Q}}
    \cr
    &
    \quad\times\bigg[\frac{1}{\omega+i\eta-E_{\bm{k},Q}}-\frac{1}{\omega+i\eta+E_{\bm{k},Q}}\bigg],
\end{align}
where $\eta$ is an infinitesimally small value, $E_{\bm{k},B(Q)}=\sqrt{\xi_{\bm{k},B(Q)}^2+|\Delta_{B(Q)}|^2}$ is the BCS quasiparticle dispersion,
and $v^2_{\bm{k},B(Q)}=1-u^2_{\bm{k},B(Q)}=\frac{1}{2}\Big(1-\frac{\xi_{\bm{k},B(Q)}}{E_{\bm{k},B(Q)}}\Big)$ is the quasiparticle weight~\cite{mahan2013many}.
To proceed the calculation, we consider the color superconducting phase in quark matter~\cite{RevModPhys.80.1455}.
For simplicity, we assume that the color superconducting phase is characterized by a single diquark gap $\Delta_Q$.
This situation might be relevant for single-flavor pairing states~\cite{iwasaki1995pairing,iwasaki1995superconductivity} or 2SC phase~\cite{alford1998qcd,PhysRevLett.81.53}.
The three-quark propagator can be decomposed into
\begin{align}
    G^{(ij){\rm ret.}}_{Q,\{\bm{k}_i,\sigma_i,\tau_i,a_i\}}(\omega)&\simeq \int\frac{d\omega_1}{2\pi}\int\frac{d\omega_2}{2\pi}\int\frac{d\omega_3}{2\pi}\cr
    &\times \delta(\omega_1+\omega_2+\omega_3-\omega)
    g^{(ij){\rm ret.}}_{Q,\bm{k}_1,\sigma_1,\tau_1,a_1}(\omega_1)
    \cr
    &\times g^{(ij){\rm ret.}}_{Q,\bm{k}_2,\sigma_2,\tau_2,a_2}(\omega_2)g^{(ij){\rm ret.}}_{Q,\bm{k}_3,\sigma_3,\tau_3,a_3}(\omega_3),\cr
\end{align}
where $i,j=1,2$.
Eventually,
in the limit of $\Delta\mu\rightarrow 0$ in 
$I_{\rm J}(t)$,
we obtain the DC Josephson current
given by
\begin{widetext}
    \begin{align}
    \label{eq:IDC}
    I_{\rm DC}=-4
    \sum_{\rm B}
    \sum_{\rm Q_1,Q_2,Q_3}
    &\int_{-\infty}^{\infty} \frac{d\omega}{2\pi}\mathcal{T}^2
    \Big[\operatorname{Re}G^{(12){\rm ret.}}_{Q,\{\bm{k}_i,\sigma_i,\tau_i,a_i\}}(\omega)\operatorname{Im}G^{(21){\rm ret.}}_{B,\bm{K},\sigma_B,\tau_B}(\omega)f(\omega)\nonumber\\
    &+\operatorname{Im}G^{(12){\rm ret.}}_{Q,\{\bm{k}_i,\sigma_i,\tau_i,a_i\}}(\omega)\operatorname{Re}G^{(21){\rm ret.}}_{B,\bm{K},\sigma_B,\tau_B}(\omega)f(\omega)\Big]\sin(\Delta\phi).
\end{align}
\end{widetext}
We note that 
$I_{\rm DC}\neq 0$ at $\Delta\mu\rightarrow 0$ is in contrast to the quasiparticle tunneling current
$I_{\rm qp}$ which vanishes in the limit of $\Delta\mu\rightarrow 0$.

\begin{figure}[t]
    \centering
    \includegraphics[width=\linewidth]{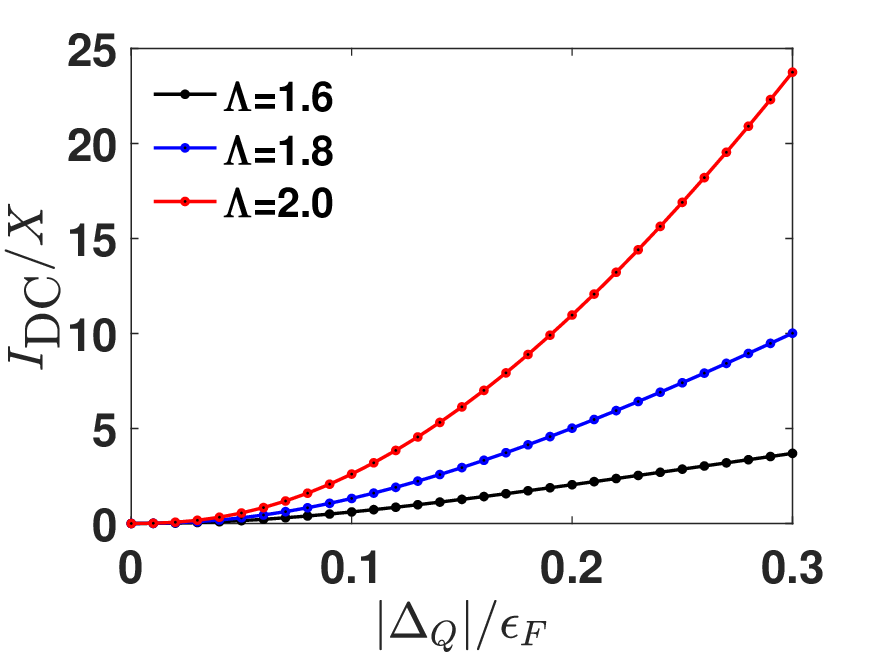}
    \caption{
    The DC Josephson current $I_{\rm DC}$ at the hadron-quark  interface for different momentum cutoff $\Lambda$
    normalized by the quark Fermi momentum $k_{\rm F}$.
  $X =N_{c}N_{s}N_f|\Delta_B|\mathcal{T}^2k_{\rm F}^{12}\sin(\Delta\phi)/(2^7\pi^{11}\epsilon^5_{\rm F})$
  is the normalization factor with color, spin and flavor degrees of freedom denoted by $N_c$, $N_{s}$, and $N_f$, respectively.
  The diquark pairing gap $|\Delta_Q|$ is taken such that it is comparable with that employed in  Ref.~\cite{PhysRevD.74.074020}.
  Since $|\Delta_B|$ is sufficiently small compared to the baryon Fermi energy around the core region, we 
  only keep the leading-order expression where $|\Delta_B|$ is absorbed into $X$ and 
  $|\Delta_B|\rightarrow 0$ is taken in the integrand of Eq.~\eqref{eq:IDC}.
    }\label{IDC}
\end{figure}

In Fig.~\ref{IDC}
we show the numerical results of DC Josephson current at a constant tunneling coupling $\mathcal{T}$, where $X=N_{c}N_{s}N_f|\Delta_B|\mathcal{T}^2k_{\rm F}^{12}\sin(\Delta\phi)/(2^7\pi^{11}\epsilon^5_{\rm F})$ is the normalizing factor with the quark Fermi energy $\epsilon_{\rm F}$ ($N_c$, $N_s$, and $N_f$ are the color, spin, and flavor degrees of freedom, respectively, relevant to each pairing channel).
$\Lambda$ is the dimensionless ultraviolet momentum cutoff normalized by the quark Fermi momentum $k_{\rm F}$.  
$I_{\rm DC}$ increases monotonically with increasing $|\Delta_{Q}|$.
In particular, one can see a non-linear $|\Delta_Q|$-dependence of $I_{\rm DC}$ even at small $|\Delta_Q|$, because the anomalous three-quark propagator $G_{Q}^{(12){\rm ret.}}$ in Eq.~\eqref{eq:IDC} is proportional to $|\Delta_Q|^3$.
This result indicates that the DC Josephson current between baryon superfluid and color superconductivity is given by
\begin{align}
    I_{\rm DC}\propto |\Delta_B||\Delta_Q|^3\sin(\phi_B-3\phi_Q),
\end{align}
in contrast to the conventional superconducting junction with $I_{\rm DC}\propto|\Delta_1||\Delta_2|\sin(\phi_1-\phi_2)$ (where $\Delta_{n=1,2}=|\Delta_{n}|e^{-i\phi_n}$ is the order parameter of two superconductors~\cite{mahan2013many}).
Moreover, the fact
that $I_{\rm DC}$ is proportional to $\sin(\Delta\phi)\equiv\sin(\phi_B-3\phi_Q)$
indicates that the phase mode in baryon and quark superfluids plays a crucial role at the interface at low temperatures. 
It is reminiscent of the vortex continuity from the hadronic to dense color superconducting phase~\cite{PhysRevD.99.036004}.

We note that our result explicitly depends on $\Lambda$ because we assumed the contact-type couplings (in other words, the momentum-independent pairing gaps), which exhibit an ultraviolet divergence.
It is known that the ultraviolet divergence in relativistic fermions with the contact coupling cannot be renormalized in the context of the NJL model studies~\cite{RevModPhys.80.1455}.
In non-relativistic fermions, the contact coupling can be renormalized by using the scattering length~\cite{strinati2018bcs,ohashi2020bcs}.
On the other hand, the higher-order low-energy scattering parameters are needed at the high-density regime to obtain the renormalized baryon-baryon interactions.
Although these details are important for further analysis,
it is out of scope in this paper to examine the tunneling transport at the hadron-quark interface.
Also, in the numerical calculation of $I_{\rm DC}$ shown in Fig.~\ref{IDC},
we practically keep a nonzero temperature given by $T/T_{\rm F}=10^{-3}$ (where $T_{\rm F}$ is the quark Fermi temperature), but confirmed that the resulting finite-temperature effect is negligible in our calculation.

\section{Summary}
\label{sec:summary}
In this paper,
using the Schwinger-Keldysh approach combined with the tunneling Hamiltonian,
we have studied tunneling transport at the interface of hadron and quark matter,
which may be relevant to the deep inside of a massive neutron star. 
To overcome the limitations of the conventional BdG approach regarding the tunneling transport involving the hadron formation, we have developed the field-theoretical approach that takes into account the hadron--multi-quark conversion microscopically.
We have shown the perturbative expression of the tunneling current across the hadron-quark interfaces.
Moreover, it is found that the friction at the hadron-quark interfaces can be studied in a similar manner.
To demonstrate our approach, we have calculated the DC Josephson current at the interface between baryon superfluid and color superconductivity,
which is found to be associated with the phase difference between two order parameters as $\Delta\phi=\phi_B-3\phi_Q$.

For future perspectives,
it is worthwhile to examine the tunneling current and the mutual frictions at interfaces under the differential rotations in a neutron star,
which may be relevant to the glitch phenomena and starquakes.
The effects of superfluid vortices and thermal transport would also be interesting topics for future work.

\begin{acknowledgments}
The authors thank
M. Sadzikowski for stimulating discussions and A. Sedrakian for useful comments.
H.~T. also thanks Daigo Oue and Mamoru Matsuo for fruitful discussions at the initial stage of this research.
This work is supported by JSPS KAKENHI for Grants (Nos.~JP22K13981, JP23K22429) from MEXT, Japan. T.Z.~is supported by the RIKEN Junior Research Associate Program.
\end{acknowledgments}

\bibliographystyle{apsrev4-2}
\bibliography{reference.bib}

\end{document}